\documentclass[twocolumn,floatfix,prl,aps,showpacs,superscriptaddress]{revtex4-1} 

\usepackage{graphicx,epsf,amsmath,amssymb}
\usepackage{dcolumn}
\usepackage{bm,bbm}
\usepackage{enumerate}
\usepackage[usenames,dvipsnames,svgnames,table]{xcolor}

\usepackage{hyperref}
\hypersetup{
    bookmarks=true,         
    unicode=false,          
    pdftoolbar=true,        
    pdfmenubar=true,        
    pdffitwindow=false,     
    pdfstartview={FitH},    
    pdftitle={Volkov-Pankratov states in topological heterojunctions},    
    pdfauthor={LPS,LPA,ENSL},     
    pdfsubject={Report},   
    pdfcreator={LPS,LPA,ENSL},   
    pdfproducer={LPS,LPA,ENSL}, 
    pdfkeywords= {electrons} {massive} {surface} 
    pdfnewwindow=true,      
    colorlinks=true,       
    linkcolor=blue,          
    citecolor=red,        
    filecolor=magenta,      
    urlcolor=cyan           
}

\begin{document}

\preprint{APS/123-QED}

\title{Volkov-Pankratov states in topological heterojunctions}

\author{S. Tchoumakov}
\affiliation{Laboratoire de Physique des Solides, CNRS UMR 8502, Univ. Paris-Sud, Universit\'e Paris-Saclay, F-91405 Orsay Cedex, France}

\author{V. Jouffrey}
\affiliation{Universit\'e de Lyon, ENS de Lyon, Universit\'e Claude Bernard, CNRS, Laboratoire de Physique, F-69342 Lyon, France}

\author{A. Inhofer}
\affiliation{Laboratoire Pierre Aigrain, D\'epartement de physique de l'ENS, Ecole normale sup\'erieure, PSL Research University, Universit\'e Paris Diderot, Sorbonne Paris Cit\'e, Sorbonne Universit\'es, UPMC Univ. Paris 06, CNRS, 75005 Paris, France}

\author{E. Bocquillon}
\affiliation{Laboratoire Pierre Aigrain, D\'epartement de physique de l'ENS, Ecole normale sup\'erieure, PSL Research University, Universit\'e Paris Diderot, Sorbonne Paris Cit\'e, Sorbonne Universit\'es, UPMC Univ. Paris 06, CNRS, 75005 Paris, France}

\author{B. Pla\c cais}
\affiliation{Laboratoire Pierre Aigrain, D\'epartement de physique de l'ENS, Ecole normale sup\'erieure, PSL Research University, Universit\'e Paris Diderot, Sorbonne Paris Cit\'e, Sorbonne Universit\'es, UPMC Univ. Paris 06, CNRS, 75005 Paris, France}

\author{D. Carpentier}
\affiliation{Universit\'e de Lyon, ENS de Lyon, Universit\'e Claude Bernard, CNRS, Laboratoire de Physique, F-69342 Lyon, France}

\author{M. O. Goerbig}
\affiliation{Laboratoire de Physique des Solides, CNRS UMR 8502, Univ. Paris-Sud, Universit\'e Paris-Saclay, F-91405 Orsay Cedex, France}

\date{\today}

\begin{abstract}
We show that a smooth interface between two insulators of opposite topological $\mathbb{Z}_2$ indices possesses multiple surface states, both massless and massive. 
While the massless surface state is non-degenerate, chiral and insensitive to the interface potential, the massive surface states only appear for a 
sufficiently smooth heterojunction. 
The surface states are particle-hole symmetric and a voltage drop reveals their intrinsic relativistic nature, similarly to Landau bands of Dirac electrons in a magnetic field. We discuss the relevance of the massive Dirac surface states in recent ARPES and transport experiments.
\end{abstract}

\pacs{Condensed matter}

\maketitle

Topological gapped phases are fascinating new states of matter. 
Their hallmark is the existence of gapless chiral states at their surface that have been probed in topological insulators by ARPES [\onlinecite{exp1,exp2,exp3}], transport [\onlinecite{exp4,exp5}] and STM [\onlinecite{stm}] measurements. 
Our common understanding of their existence lies in the necessary gap closing at the interface between two insulators characterized by different 
topological invariants [\onlinecite{reviewtopo, interpret, tbc}]. 
Moreover, the same experiments that detect these gapless states also evidence multiple massive surface states in both the conduction and valence bands. 
These additional surface states have been attributed to conventional properties of the interface, such as band bending, unrelated to the topological nature of the insulators. 
Indeed, previous studies consider that the amplitude of band bending is strong enough to confine states in both the conduction and the valence bands  [\onlinecite{exp1,exp2,th1}]. 
From this point of view, one would consider the two types of surface states to be of different origin: topological for the massless states and due to strong band bending for the massive ones. 

Here we show that this is not necessarily the case. 
We describe the interface between a topological and a normal insulator or vacuum, which is a \emph{topological heterojunction} (THJ), within a model where the gap inversion occurs over a finite interface size $\ell$, and we show that it hosts multiple surface states, both massless and massive.
In the limit of a wide interface, where $\ell$ is much larger than an intrinsic material-dependent length scale $\xi$, these states are similar to Landau bands of a Dirac material in a magnetic field. 
The gapless surface state is then analogous to the $n = 0$ Landau band and it only depends on the properties far away from the interface as in the Aharonov-Casher argument \cite{acasher,kawaraba}. 
Furthermore, as for Landau bands, the massive surface states appear in both the conduction and the valence band and are sensitive to the details of the interface.
Prior to the recent discussion within the context of topological insulators \cite{tbc}, massless and massive surface states at an interface with gap inversion had been studied theoretically back in the 1980s by Volkov and Pankratov in a set of pioneering papers \cite{volk, volk1, volk2}, albeit in the simplified framework of a symmetric interface. We thus call such massive surface states \emph{Volkov-Pankratov states} (VPS).
In the present paper we stress the topological origin of this type of confinement which is tightly related to relativistic physics and inherently different from 
confinement in a conventional quantum well. 
Moreover, we characterize the properties of the surface states as a function of various properties of the THJ, such as a gap asymmetry between the two materials and surface band bending. 

We consider the interface between two semiconductors with inverted, $\Delta_1 <0$, and conventional, $\Delta_2 >0$, gaps in the simplest situation where both gaps are located at a single point of the Brillouin zone, e.g. the $\Gamma$ point.  
Hence we model each insulating phase with the generic $k \cdot P$ Hamiltonian around the $\Gamma$ point, describing in particular Bi$_2$Se$_3$ \cite{tbc},
\begin{multline}\label{eq:hamiltonian}
	\hat{H}_0 = \mu ~\hat{\mathbbm{1}} \otimes \hat{\mathbbm{1}} + \Delta ~\hat{\mathbbm{1}} \otimes \hat{\tau}_z 
	+ v_F k_z ~ \hat{\mathbbm{1}}\otimes\hat{\tau}_y 
	\\
	+ v_F \left( k_y \hat{\sigma}_x - k_x \hat{\sigma}_y \right) \otimes \hat{\tau}_x
\end{multline}
where we assume $v_F>0$ and set $\hbar = 1$ hereafter. 
The $\hat{\sigma}$ and $\hat{\tau}$ Pauli matrices act respectively on spin and orbital subspaces.
In a first place, $\mu$ is set to zero in both semi-conductors. The spectrum of \eqref{eq:hamiltonian} consists of two doubly degenerate bands $\varepsilon_{\bf k}^{(\pm)} = \pm \sqrt{\Delta^2 + v_F^2 k^2}$ 
and the two insulators only differ by their band gaps $\Delta_1$ and $\Delta_2$.

\begin{figure*}[th]
	\centering
	\includegraphics[width=\textwidth]{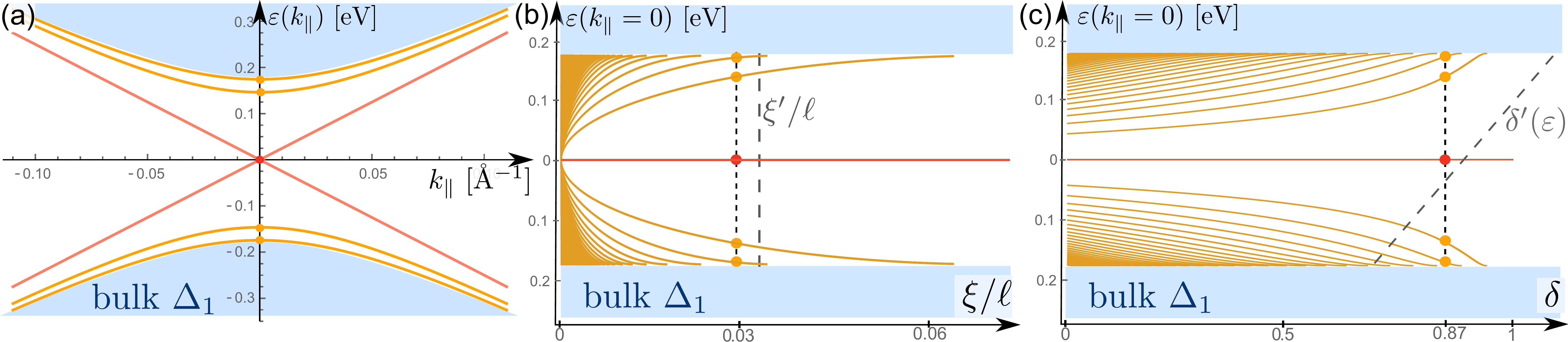}
	\caption{(a) Surface states obtained for $2\Delta_1 = -0.35$ eV, $2\Delta_2 = 5$ eV, $v_F = 2.5$ eV\AA~ and $\ell = 6$ nm. The corresponding reduced parameters are $\delta = 0.87$ and $\xi/\ell = 0.03$. 
	We show the band dispersion of the gapless surface state (in red), the massive surface states (in orange), and 
	the bulk states (in blue). 
	(b,c) Gaps of the surface states (in orange) for $v_F = 2.5$ eV\AA~, $2\Delta_1 = -0.35$ eV and (b) as a function of the sharpness of the interface $\xi/\ell$ with $\delta = 0.87$
	and (c) as a function of the gap asymmetry $\delta$ with $\xi/\ell = 0.03$. An applied electric field renormalizes the values of $\xi$ and $\delta$ (gray dashed lines), as a function of energy for $\delta$.}
	\label{fig:spectra}
\end{figure*}

We model the interface between the two insulators, chosen as normal to the $z$-direction, by a $z$-dependent $k\cdot P$ 
Hamiltonian, $\hat{H}_s$, which smoothly interpolates between the two bulk Hamiltonians over a characteristic size $\ell$. 
This amounts to replacing $\Delta$ in \eqref{eq:hamiltonian} with a smoothly interpolating gap, $\Delta(z)$, such that 
$\Delta (z\to-\infty) = \Delta_{1}$ and $\Delta (z\to +\infty) = \Delta_{2}$. 
In addition to the interface width $\ell$, one finds a 
natural length $\xi = 2 v_F/|\Delta_1-\Delta_2|$, and for a smooth interface one has $\ell > \xi$, while for a sharp interface $\ell < \xi$. 
The interface width $\ell$ depends on the material and growth technique but we expect that a smooth interface generically occurs between a small gap topological insulator such as Bi$_2$Se$_3$ ($2\Delta_1 = - 0.35$ eV \cite{exp1}, $v_F = 2.5$ eV\AA \cite{vf1}) and a large gap insulator such as HfO$_2$ ($2\Delta_2 = 5$ eV) since then $ \xi \sim 2$ \AA ~is rather small. 
On the other hand, for the interface between two small gap insulators, with e.g. $2\Delta \sim 0.25$ eV, we estimate $\xi \sim 2$ nm and thus, depending on $\ell$, there can be either a smooth or a sharp 
interface \footnote{The width of the interface can in principle be derived within a mean-field approximation by choosing $\ell$ that minimizes the energy as in Ref. \cite{ssh}.}.  

A convenient choice to describe the $z$-dependent gap around the THJ is $\Delta (z) = \frac12 (\Delta_2 - \Delta_1)[\delta + \tanh(z/\ell)]$, where we have introduced 
the {\it gap asymmetry} $\delta = (\Delta_1 + \Delta_2 ) / (\Delta_2 - \Delta_1)$. The situation with opposite gaps, $\delta = 0$, for which the Schr\"odinger equation 
resembles to the P\"{o}schl-Teller equation \cite{poteller}, was studied in \cite{volk, volk1, volk2}. Here, we solve the more relevant situation with $\delta \neq 0$ 
with the use of hypergeometric functions \cite{appsol}. This yields the spectrum of surface states 
\begin{equation}\label{eq:spectanh_simpler}
	\varepsilon_{n,\pm}(k_x,k_y) = \pm v_F
	\sqrt{ k_x^2 + k_y^2 +  1/\ell_n^{2}  },
\end{equation}
where the characteristic length $\ell_n$ of each mode in the $z$-direction is
\begin{equation}\label{eq:ln}
 	\frac{1}{\ell_n^{2}}  =  \frac{2n}{\ell \xi}  
	\left( 1 - n \frac{\xi}{2 \ell }  \right)
	\left[ 1 - \left( \frac{\delta}{1 - n \frac{\xi }{\ell}} \right)^2 \right], 
\end{equation}
which depends on the integer values $n$ delimited by
\begin{align}\label{eq:Nmax}
	n < N = \frac{\ell }{ \xi}  \left( 1 - \sqrt{\left|\delta \right|} \right).
\end{align}
The corresponding band dispersions are illustrated in Fig.~\ref{fig:spectra}(a) with the $n = 0$ state in red and the $n \geq 1$ states in orange. 
The inequality (\ref{eq:Nmax}) yields surface states only if $|\delta| < 1$, i.e. if the two semiconductors have gaps of opposite sign. 
Moreover, as illustrated in Fig.~\ref{fig:spectra}(b), the number of states depends on the THJ geometry with many massive ($N \geq 1$) hole- \emph{and} electron-like surface states for a smooth interface but with only the single massless $n=0$ surface state for a sharp interface.
This kind of quantization is reminiscent of that in a uniform magnetic field which happens to be analogue to the limit of a linearized potential for $\ell \gg \xi$ which we discuss in \cite{appsol} and which was already discussed in the context of Majorana surface states in \cite{majo0,majo1,majo2}.
Note that the occurrence of the additional VPS still fulfills the topological $\mathbb{Z}_{2}$ constraint \cite{z2} since the $n = 0$ state is simply degenerate while the VPS ($n \geq 1$) are doubly degenerate, such that the parity of the number of surface states is unchanged. 
We stress that the existence of both the massless and the massive surface states requires the heterojunction to be topological, $\Delta_1 \Delta_2 < 0$, in contrast to standard band bending massive surface states. 
     
The peculiar nature of the VPS is further illustrated by their response to an electric field applied perpendicular to the interface. 
We expect the number of VPS, expressed in Eq. \eqref{eq:Nmax}, to be small since their band gap has to be smaller than the smallest bulk band gap, $\textrm{min}(|\Delta_1| , |\Delta_2|)$, and that a large gap asymmetry $\delta$ decreases their number as illustrated in Fig.~\ref{fig:spectra}(c). 
However, we show below that the visibility of VPS can be strongly enhanced by an electric field, possibly originating from surface band bending due to chemical doping or an applied gate voltage under Dirac screening \cite{screening}.  
\begin{figure*}[thb]
	\centering
	\includegraphics[width=\textwidth]{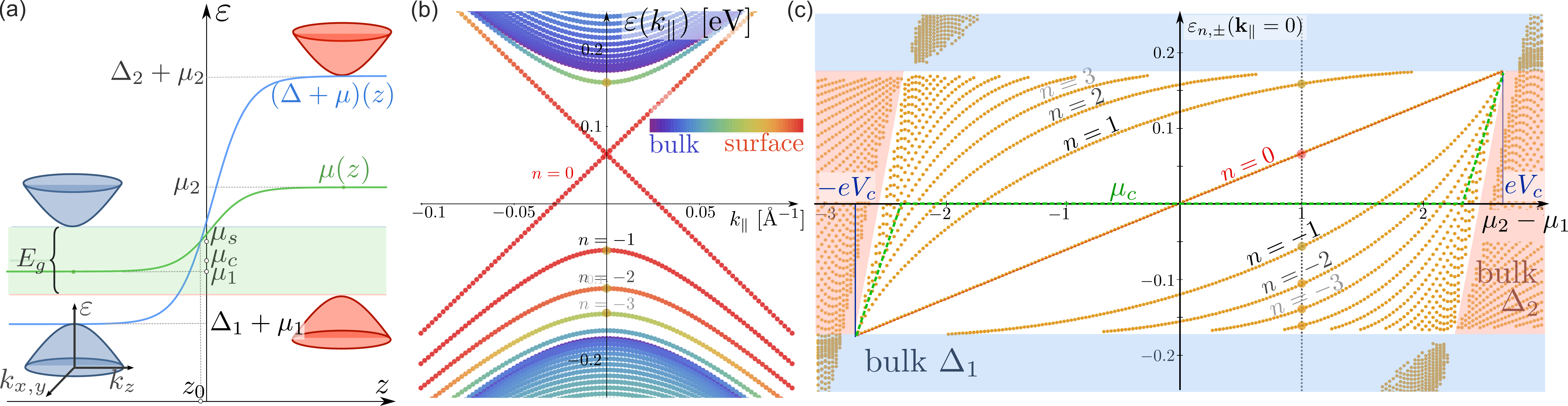}
	\caption{Surface states of a THJ with $n-$ and $p-$doped insulators. (a) Sketch of the heterojunction with a $\mu_2 - \mu_1$ drop in the chemical potential. The pn-junction is characterized by a chemical potential $\mu_c$ and a band gap common 
	to both insulators (shaded green region) of size $E_g = |\Delta_2 - \Delta_1 - (\mu_2 - \mu_1)|$ which vanishes for a large chemical potential or voltage drop, defining a breakdown voltage. 
	The surface states are centered around the position $z_0$ where the gap vanishes, with surface chemical potential $\mu_S \approx \mu(z_0) = \frac12(\mu_1 + \mu_2)-\frac12 (\mu_2-\mu_1)\delta$. 
	(b) Spectrum obtained by numerical simulation with the same parameters as in Fig. \ref{fig:spectra},  $f(z/\ell) = \tanh(z/\ell)$ and centered around the band gap of the inverted insulator (in blue). 
	We use $\mu_2 - \mu_1 = 1$ eV, and the colors represent the amplitude of surface density $\rho_S \approx |\Psi(z=z_0)|^2$ for each state. 
	(c) Energy $\varepsilon_{n,\pm}({\bf k}_{\parallel} = 0)$ of the surface states as a function of the chemical-potential difference $\mu_2-\mu_1$. Multiple VPS appear in the valence band for $|\mu_2-\mu_1|$ closer to $|eV_c| = |\Delta_2-\Delta_1| = 2.675$ eV for a positive and strong gap asymmetry, $\delta = 0.87 \sim 1 = |\delta|_{\max}$. 
	}
	\label{fig:efield}
\end{figure*}
We consider a chemical potential $\mu$ in Eq.~\eqref{eq:hamiltonian} that varies over the interface with the same profile $\mu(z) = \frac12 (\mu_2 - \mu_1)f(z/\ell)$ as the gap $\Delta(z)$, as depicted in Fig.~\ref{fig:efield}(a), and with opposite potentials in each semiconductor.
We account for this potential with a Lorentz boost \cite{applorentz} which shows that the surface states only exist if the relativistic parameter 
$\beta = -(\mu_2 - \mu_1) /(\Delta_2 - \Delta_1) \in [-1,1]$. This reminds the behavior of Landau levels in crossed electric and magnetic fields \cite{lukose} with a typical electric field $\mathcal{E} = (\mu_2-\mu_1)/e\ell$ that has to be smaller to a \emph{critical electric field} $\mathcal{E}_c = (\Delta_2-\Delta_1)/e\ell$, corresponding to a breakdown voltage $V_c = (\Delta_2-\Delta_1)/e$, above which the surface states disappear. 
This critical behavior happens because the surface states exist within the band gap common to both semiconductors depicted in Fig. \ref{fig:efield}(a) and that it vanishes above the breakdown voltage.
For voltage drops below $V_c$, the Lorentz boost renormalizes the parameters in the Schr\"odinger equation $\hat{H}^{\prime}_{s} |\Psi^{\prime}\rangle = \varepsilon|\Psi^{\prime}\rangle$ \cite{applorentz}. The interface is sharpened by an increase in the intrinsic length as $\xi^{\prime}/\xi = 1/\sqrt{1-\beta^2}$, while the effective gap asymmetry becomes 
\begin{align}
    \delta' = \frac{1}{1-\beta^2}\left[ \delta + \frac{\varepsilon(\mu_2-\mu_1)/2}{(v_F/\xi)^2} \right],
\end{align}
and the effective Fermi velocity decreases as $v_F^{\prime}/v_F  = \sqrt{1 - \beta^2}$. 
These renormalized parameters decrease the VPS band gaps \cite{applorentz} and alter their spectrum in a way that depends on the sign of the voltage drop $\mu_2-\mu_1$, relative to the gap asymmetry $\delta$. As shown in Fig.~\ref{fig:spectra}(c), the renormalized $\delta^{\prime}$ is such that if $\mu_2-\mu_1 $ and $\Delta_2 - \Delta_1$ have the same signs, then increasing the voltage drop shifts up the spectrum of VPS and increases the number of particle-like VPS. This is the situation depicted in Fig.~\ref{fig:efield}(c).
Conversely, if $\mu_2-\mu_1 $ and $\Delta_2 - \Delta_1$ have opposite signs, an increasing electric field shifts down the VPS spectrum and increases the number of hole-like VPS. 
This unique behavior is a hallmark of VPS in a THJ, revealing their intrinsic relativistic origin. 
The spectrum of surface states has to be obtained self-consistently if the original spectrum depends explicitly on $\delta$, as in Eq.~\eqref{eq:spectanh_simpler}. This can be avoided in simplified models \cite{exp4,applorentz}.
Alternatively, we can directly solve a discretized version of Eq. \eqref{eq:hamiltonian} for a generic $\tanh (z/\ell)$ interface. The results are shown in Fig. \ref{fig:efield}(b-c) and confirm the unique effect of a voltage drop on the VPS of a THJ.

\begin{figure}[thb]
	\centering
	\includegraphics[width=\columnwidth]{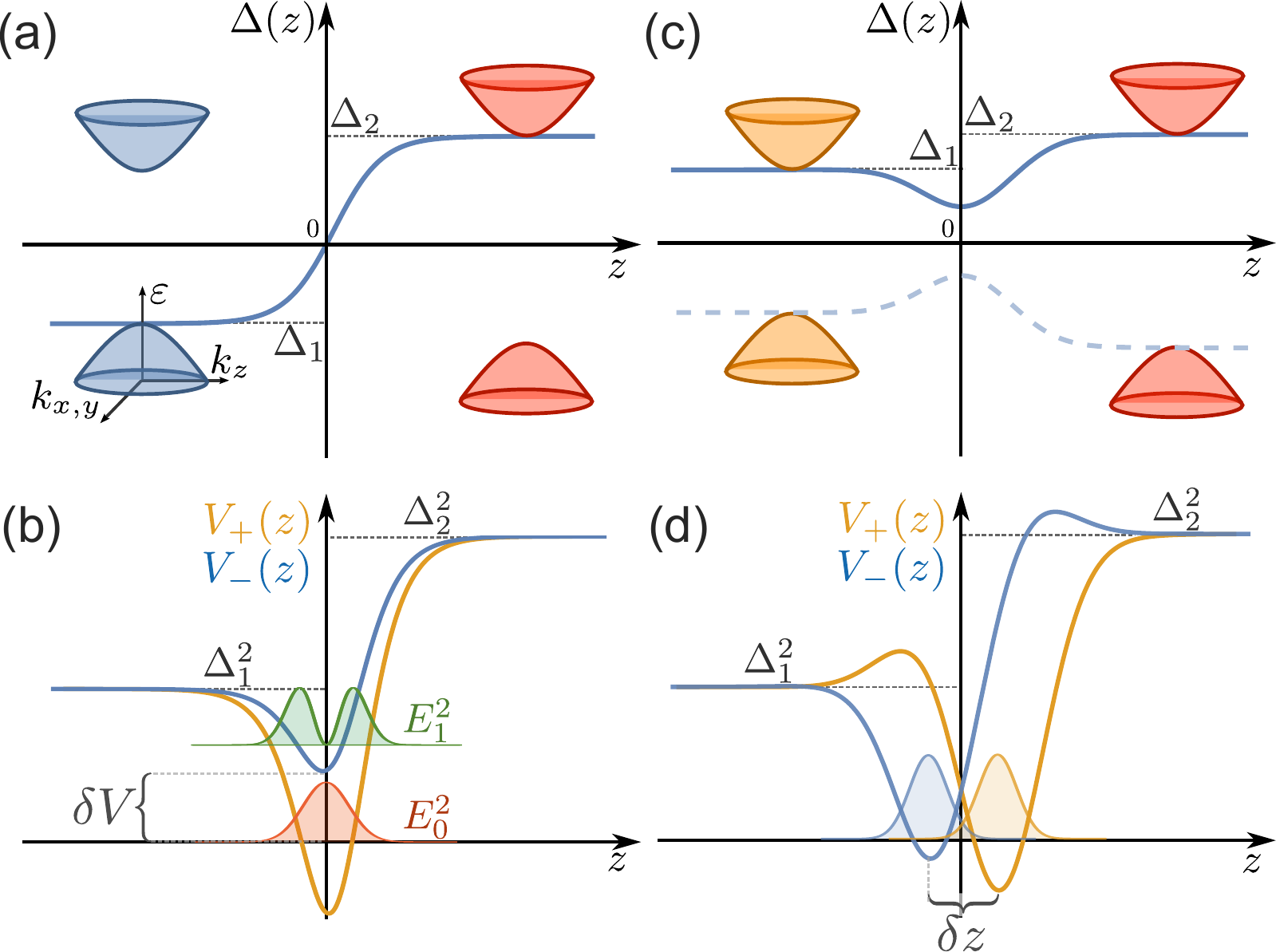}
	\caption{Sketch of the position-dependent gap $\Delta(z)$ and the corresponding confining potentials $V_{\pm}(z)$ of the wavefunction components
	$\Psi_{\pm}$. In (a),(b) is the case of band inversion (THJ) and in (c),(d) is the case of a gap minima $|\Delta(z)|$ in a conventional heterojunction. 
    In (b) The potentials minima are shifted in energy by $\delta V \sim v_F^2/\ell\xi$ and in (d) the potential minima are shifted in position by $\delta z \sim v_F/2\Delta_{\min}$. 
	It is only in the case of a THJ that one finds a non-degenerate bound state for the $V_-$ potential.}
	\label{fig:interface}
\end{figure}

Let us comment on the intrinsic difference in nature between VPS in a THJ and conventional nonrelativistic states localized in a potential well at an interface. 
In both cases, the interface is described by a change in the gap $\Delta(z)$ for which the parity symmetry $\mathcal{P} = \hat{\sigma}_z\otimes\hat{\tau}_z$ leads to a particle-hole symmetric spectrum which is absent in the case of electrostatic confinement.
Moreover, a general surface state $\Psi = (\phi_{1,+},\phi_{2,+},\phi_{1,-},\phi_{2,-})$ written over the eigenbasis of the parity operator $\mathcal{P}$ as in \cite{appsol} can be determined by squaring the time-independent Schr\"odinger equation associated with Eq.~\eqref{eq:hamiltonian}. In doing so, we obtain for each component $\phi_{i,s}$ ($i = 1,2$, $s = \pm$) a one-dimensional nonrelativistic Schr\"odinger equation
\begin{equation}\label{eq:masspot}
	\left[-  v_F^2 \partial_z^2 + V_{s}(z) \right] \phi_{i,s} =   (\varepsilon^2 - v_F^2 k_{\parallel}^2) \phi_{i,s} ,
\end{equation}
with a potential $V_{\pm}(z) = \Delta(z)^2  \mp v_F  \partial_z \Delta(z)$, and associated eigenvalues $E_n^2 = (\varepsilon^2 - v_F^2 k_{\parallel}^2)$.
The chiral symmetry relates the two solutions associated with both $V_{\pm}(z)$ potentials {\it via} the relation $\phi_{i,s} = -[v_F \partial_z - s \Delta(z)]\phi_{i,-s}$. 
Thus every state is doubly degenerate except for the $n = 0$ surface state which is analyzed in detail below and which we call the \emph{chiral surface states} (CSS). 
These CSS are the topologically protected ones, which one obtains based on topological invariants and that survive in the limit where the interface becomes sharp, $\ell/\xi \rightarrow 0$. 
This means that the CSS are completely independent of the $\Delta(z)$ profile as shown by the Aharonov-Casher argument in \cite{appacasher}. 
In Fig.~\ref{fig:interface}, we represent two situations for this Schr\"odinger equation : a THJ in Fig.~\ref{fig:interface}~(a,b) where $\Delta(z=0) = 0$ and
a conventional heterojunction in Fig.~\ref{fig:interface}~(c,d) where there is no band inversion but a local minimum of the band gap $|\Delta(z)|$. 

The physical properties of these surface states can be captured by an analysis of Eq.~\eqref{eq:masspot}. 
In the THJ case, $V_+(z)$ necessarily has a minimum regardless of the interface size because $\Delta(z=0)=0$ and thus 
$V_+(z=0)= -v_F\partial_z\Delta(z=0)<0$. 
The associated bound state is then the CSS with $E_0=0$ whose nondegenerate nature manifests itself here by the fact that $V_-(z=0)= +v_F\partial_z\Delta(z=0)>0$, precluding another $E=0$ state. 
The existence of VPS necessitates that both $V_{-}(z)$ and $V_{+}(z)$ have a minimum because of the chiral symmetry, in particular this necessitates $V_-(0)<\min\{\Delta_1^2,\Delta_2^2\}$. 
One can estimate $V(z \sim 0)\sim v_F\partial_z\Delta(z=0)\sim v_F^2/\ell \xi$, such that the condition for appearance of  
VPS reads $v_F^2/\ell \xi <\min\{\Delta_1^2,\Delta_2^2\} = v_F^2(1 - |\delta|)^2/\xi^2$, which agrees with Eq. (\ref{eq:Nmax}) obtained for a particular choice of $\Delta(z)$ discussed above. 
This discussion can also be extended to a conventional heterojunction with no gap inversion, as studied in \cite{doublesqw,exp5}. 
We expect bound states if the gap function $|\Delta(z)|$ has a minimum $|\Delta_{\rm min}|<\min\{|\Delta_1|,|\Delta_2|\}$ as depicted in Fig.~\ref{fig:interface}(c).
Indeed Eq. (\ref{eq:masspot}) provides a pair of VPS shown in Fig.~\ref{fig:interface}(d) since both potentials satisfy $V_{\pm,\min} \simeq \Delta_{\rm min}^2 <\min\{\Delta_1^2,\Delta_2^2\}$. 
In this case, all massive surface states are degenerate with a linear combinations of $|\pm\rangle$ states located on either side of the interface with a separation $\delta z \simeq v_F/2\Delta_{\min}$, similarly to the case of a thin confined insulator recently discussed in \cite{exp5}.

The above description of a THJ shows the existence of massive states intimately related to the relativistic and topological nature of the interface. While a clear determination 
of the nature of these surface states requires to study their dependence on an external electric field, along the lines of Ref.~\onlinecite{exp4} for strained bulk HgTe, 
their occurrence in previous ARPES experiments in Bi$_2$Se$_3$ and Bi$_2$Te$_3$ aged in an oxidizing atmosphere \cite{exp2,exp3} can be critically discussed within our theory.
While electrostatic band bending can reproduce the electron-like surface states, the hole-like ones necessitate an associated energy scale larger than the bulk valence band width, unlikely in particular in Bi$_2$Te$_3$ \cite{exp3}. On the other hand in our theoretical description both electron- and hole-like VPS arise on equal footing due to the underlying particle-hole symmetry in a THJ. 
In the experiments reported in \cite{exp2,exp3}, one observes 2 or 3 electron- and hole-like surface states. 
For Bi$_2$Se$_3$, published values for $v_F = 2.3\dots 5$ eV\,\AA\ \cite{vf1,vf2} and $2\Delta = 350$ meV \cite{exp1} yield a characteristic interface width 
$\xi \approx v_F/\Delta = 6.5\dots 23$ \AA. 
In an oxidizing atmosphere, the depth of the oxide layer can be estimated as $\ell \approx 10\dots 20$ \AA~ \cite{oxlayer}, which leads to an expected number of  
$N \approx \ell/\xi = 1\dots 3$ VPS, in agreements with experiments \cite{exp2,exp3}. Moreover, we find $\ell_S = \sqrt{\ell\xi} \approx 8\dots 20$\AA ~ and thus 
VPS energy gaps $\Delta_{VPS} \approx v_F/\ell_S = 100\dots 600$ meV in reasonable agreement with \cite{exp2,exp3}.
In our theory we expect similar band gaps for the electron and hole-like VPS which is observed experimentally in Bi$_2$Se$_3$ as we discuss in \cite{appom}. This is a strong indication of the topological origin of the surface states. 
The ARPES measurement of both the band gap and the Fermi velocity of VPS as a function of band bending could help to get better insight into their relativistic nature.
Moreover we expect the absence of VPS for In-doped Bi$_2$Se$_3$ in an oxidizing atmosphere for a doping that turns the material into a normal semiconductor and which has been recently discussed in \cite{indoped}.

In conclusion we have shown that THJs, characterized by a continuous gap inversion between two semiconductors, generically host massive surface states in 
addition to the usual massless chiral state. 
A condition for the existence of these VPS is a sufficiently smooth interface $\ell > \xi$ as compared to the material-dependent characteristic length $\xi=2v_F/|\Delta_1-\Delta_2|$ and a small gap asymmetry $\delta = (\Delta_1 + \Delta_2 ) / (\Delta_2 - \Delta_1)$. 
Both the massless and the massive surface states are intrinsically relativistic, revealed in the particle-hole symmetry of their spectrum as well as by the effect of electrostatic band bending. 
The surface states only persist below the breakdown voltage but the number of electron- or hole-like surface states can be increased by band bending depending on the sign of the voltage drop with respect to the direction of increase in the gap. 
These surface states are reminiscent of Landau bands, in particular for a wide THJ ($\ell\gg \xi$). 
The existence of such VPS and their relativistic nature is relevant for recent ARPES measurements on aged Bi$_2$Se$_3$ and Bi$_2$Te$_3$ \cite{exp1,exp2,exp3} and have been identified in transport measurements on strained bulk HgTe \cite{exp4}. 
Uncovering the signature of these states and their electric field dependence for other probes such as in tunneling spectroscopy is a natural and exciting perspective.

We would like to thank M. Civelli, B. A. Assaf and C. Quay for fruitful discussions.

\bibliographystyle{apsrev4-1}
\bibliography{bibliography_SurfaceState}

\newpage
\begin{widetext}
\begin{center}
	{\bf Appendix}
\end{center}

\section{I. Surface states of a THJ}

Since $k_z$ does not commute with position, $[z,k_z] = i$, we simplify the Hamiltonian in Eq.~(1) of the main text with the following rotation 
$|\Psi\rangle = e^{-i \pi \hat{\mathbbm{1}} \otimes \hat{\tau}_y/4}|\Psi'\rangle$ to the {\it chiral} basis (with $\mu=0$)

\begin{align}\label{eq:Hsf}
   \hat{H}_c 
   = 
   \left[
       \begin{array}{cccc}
           0 & v_F k_+ & 0 & \hat{a}\\
           v_F k_- & 0 & \hat{a} & 0\\
           0 & \hat{a}^{\dagger} & 0 & -v_F k_+\\
           \hat{a}^{\dagger} & 0 & -v_F k_- & 0
       \end{array}
   \right],
\end{align}
where we introduce $k_{\pm} = k_y \pm i k_x$ and the operator $\hat{a} = -[v_F ik_z + \Delta(z)]$. In the following we solve this Hamiltonian for two different potentials $f(z/\ell)$ in $\Delta(z) = \bar{\Delta} + f(z/\ell)\delta\Delta$ with $\bar{\Delta} =\frac12(\Delta_1 + \Delta_2)$ and $\delta\Delta = \frac12 (\Delta_2-\Delta_1)$.

\subsection{a. Domain wall potential}
\label{sec:tanhcal} 
We consider the smooth domain wall $f(z) = \tanh(z/\ell)$ and search for the bound states of the Schrodinger equation. This situation is very similar 
to that in Ref. [\onlinecite{tmasstanh,TanhDW,ArakiDW}] that we will follow closely. We first perform the change of variable $s = [1-\tanh(z/\ell)]/2$ 
so that
\begin{align}
	&\hat{a} = \frac{2 v_F}{\ell} s(1-s)\partial_s - (\bar{\Delta} + \delta \Delta - 2 \delta \Delta s),\\
	&\hat{a}^{\dagger} = -\frac{2 v_F}{\ell} s(1-s)\partial_s - (\bar{\Delta} + \delta \Delta - 2 \delta \Delta s).
\end{align}
We now consider the equations for the squared Hamiltonian in Eq.(\ref{eq:Hsf}) and decompose the wavefunction into two spinors $\phi_{\sigma}$, $\sigma = \pm$ so that $\Psi = (\phi_+, \phi_-)$. One finds
\begin{align}
	&\left[\frac{1}{2} \left( \left\{ \hat{a}^{\dagger}, \hat{a} \right\} + \sigma \left[ \hat{a}^{\dagger}, \hat{a} \right] \right) - 
	(E^2 - v_F^2 k_{\parallel}^2)\right]\phi_{\sigma} = 0\\
	\implies & 
	\left[ s(1-s) \partial_s^2 + (1-2s) \partial_s - \left( \frac{\ell}{2 v_F} \right)^2 \left\{  \frac{[\bar{\Delta} + (1-2s)\delta \Delta]^2 - 
	(E^2 - v_F^2 k_{\parallel}^2)}{s(1-s)} - \sigma \frac{4 v_F \delta \Delta}{\ell} \right\} \right]\phi_{\sigma} = 0.
\end{align}
We then perform the following replacement of the wavefunction $\phi_{\sigma}(s) = s^{\alpha}(1-s)^{\beta} u_{\sigma}(s)$ in order to get 
rid of the $1/s(1-s)$ singularity and recognize Euler's hypergeometric equation [\onlinecite{Whittaker:1950}]: 
$s(1-s) \partial_s^2\phi + \left[c - (1 + a + b)s \right] \partial_s\phi - ab\phi = 0$. The parameters $\alpha$ and $\beta$ fulfill 
\begin{align}
	\left\{
		\begin{array}{l}
			\alpha^2 = \left( {\ell}/{2 v_F} \right)^2 \left[ \Delta_1^2 - (E^2 - v_F^2 k_{\parallel}^2) \right]\\
			\beta^2 = \left( {\ell}/{2 v_F} \right)^2 \left[ \Delta_2^2 - (E^2 - v_F^2 k_{\parallel}^2) \right]
		\end{array}
	\right. ,
	\label{eq:ab}
\end{align}
and in the following we choose the positive roots of those equations. The equation is now
\begin{align}
	&\left[ s(1-s) \partial_s^2 + \left[ 1 + 2\alpha - 2(1 + \alpha + \beta)s \right] \partial_s - \left\{  \left(\alpha + \beta\right)\left( \alpha + \beta + 1 \right) - \frac{\ell \delta \Delta}{v_F}\left( \sigma + \frac{\ell \delta \Delta}{v_F}\right) \right\} \right]u_{\sigma}(s) = 0,
\end{align}
which corresponds to the Euler hypergeometric differential equation. We now introduce the auxiliary parameters
\begin{align}
	\left\{
		\begin{array}{l}
			a_{\sigma} = 1/2 + \alpha + \beta + \left| 1/2  + \sigma \frac{\ell \delta \Delta}{v_F}\right|,\\
			b_{\sigma} = 1/2 + \alpha + \beta - \left| 1/2  + \sigma \frac{\ell \delta \Delta}{v_F}\right|,\\
			c = 1 + 2 \alpha.
		\end{array}
	\right.
\end{align}
For each value of $\sigma$, there are two solutions to this equation that are described by the hypergeometric 
functions [\onlinecite{Whittaker:1950}] $~_2F_1(a,b,c;s) = \sum_{n=0}^{\infty} \frac{(a)_n (b)_n}{(c)_n} z^n/n!$ with $(x)_n = x(x+1)\cdots(x+n-1)$ . 
The solutions are $u_{I}(s) = ~_2F_1(a,b,c;s)$ and $u_{II}(s) = s^{1-c} ~_2F_1(1 + a - c, 1 + b - c, 2 - c;s)$ but $u_{II}(s)$ does not 
describe bound states since $\phi_{II}(s \sim 0) = s^{\alpha}(1-s)^{\beta} u_{II}(s \sim 0) \sim s^{1 + \alpha - c}$ which diverges 
at $s = 0$ ($x = \infty$) since $1 + \alpha - c = - \alpha < 0$. On the other hand, 
while $\phi_{I}(s \sim 0) = s^{\alpha}(1-s)^{\beta}u_{I}(s \sim 0) \sim s^{\alpha}$ goes to zero at $s = 0$ ($x = \infty$), one can check its 
behavior at $s = 1$ ($x = -\infty$). We use the following relation from Ref. [\onlinecite{Whittaker:1950}] 
\begin{align}
	~_2F_1(a,b,c;s) = &\frac{\Gamma(c) \Gamma(c - a - b)}{\Gamma(c-a)\Gamma(c-b)} ~_2F_1(a,b, a + b + 1 - c; 1 - s) \\
	&+ \frac{\Gamma(c) \Gamma(a + b - c)}{\Gamma(a) \Gamma(b)} (1-s)^{c-a-b} ~_2F_1(c - a, c - b, 1 + c - a - b; 1 - s),\nonumber
\end{align}
and find that for $s \sim 1$ ($x = -\infty$)
\begin{align}
	\phi_{\sigma}(s \sim 1) \sim \frac{\Gamma(c) \Gamma(c - a_{\sigma} - b_{\sigma})}{\Gamma(c-a_{\sigma})\Gamma(c-b_{\sigma})} ( 1 - s)^{\beta} + \frac{\Gamma(c) \Gamma(a_{\sigma} + b_{\sigma} - c)}{\Gamma(a_{\sigma}) \Gamma(b_{\sigma})} ( 1 - s)^{-\beta}
\end{align}
which should diverge since $\beta > 0$ unless $\Gamma(a_{\sigma})$ or $\Gamma(b_{\sigma})$ diverges. This happens if either $a_{\sigma}$ or $b_{\sigma}$ is a negative integer and since $a > b$, this should happen on $b$. Then one has the following quantization, with $n \in \mathbb{N}$,
\begin{align}
	\sqrt{\Delta_1^2 - (E^2 - v_F^2 k_{\parallel}^2)} + \sqrt{\Delta_2^2 - (E^2 - v_F^2 k_{\parallel}^2)} = \left| \frac{2v_F}{\ell} \right| \left[ \left|\frac{1}{2} + \sigma \frac{\ell\delta\Delta}{v_F}\right| - (n + \frac{1}{2}) \right] \equiv g_{\sigma}(n),
\end{align}
and the eigenenergies are thus
\begin{align}
	E = \pm \sqrt{ v_F^2 k_{\parallel}^2 - \frac{1}{4 g_{\sigma}^2(n)}[g_{\sigma}^2(n) - 4\delta \Delta^2][g_{\sigma}^2(n) - 4\bar{\Delta}^2]}.
\end{align}
Also, as we have shown, the long range behavior of the wavefunctions is described by 
(i) $\phi_{<} = (1-s)^{\beta} = 1/(1+e^{-z/\ell})^{\beta}$ for $s \sim 1$ ($x \sim -\infty$), and 
(ii) $\phi_{>}(s) = s^{\alpha} = 1/(1+e^{z/\ell})^{\alpha}$ for $s \sim 0$ ($x \sim \infty$). This implies that $\alpha, \beta > 0$ which according to Eq. (\ref{eq:ab}) means that $E^2 - (v_F k_{\parallel})^2 < {\rm min}(\Delta_1^2, \Delta_2^2)$. Moreover, since $\alpha, \beta$ also appear in the definition of $b_{\sigma} = - n$ we find that $\alpha^2 - \beta^2 = (\ell/2v_F)^24\bar{\Delta}\delta\Delta$ and $\alpha + \beta = g_{\sigma}(n)\ell/2v_F$.
\begin{align}\label{eq:localisation}
	g_{\sigma}(n) > 2\sqrt{|\bar{\Delta} \delta \Delta|}.
\end{align}
Here, we are not interested in the exact form of the wavefunctions, and we will therefore not detail the exact form of the solutions; 
these technical details can be found in Ref. [\onlinecite{tmasstanh,ArakiDW}]. In the main text, we discuss the two limits $\ell \gg v_F/\delta \Delta$ (smooth/thick interface) and $\ell \ll v_F/\delta \Delta$ (abrupt/thin interface). In fact we can consider two situations :

\paragraph{Thick interface, $|\ell| > |v_F/2\delta\Delta|$.} In this situation, one can write 
\begin{align}
	g_{\sigma}(n) =  2\left|\delta\Delta\right| - \left| \frac{2v_F}{\ell} \right|\left[n + \frac{1-{\rm sgn}(\sigma \ell \delta\Delta/v_F)}{2}\right].
\end{align}
One finds that the two set of states $\sigma = \pm$ are thus related by a family of states with $\phi_{+} \sim |n\rangle$ and $\phi_- \sim |n + {\rm sgn}(\ell \delta \Delta/v_F) \rangle$ as in Landau levels. This implies the existence of a chiral $n = 0$ state which has the polarization $\phi_{\sigma}$ with $\sigma = -{\rm sgn}(\ell \delta \Delta/v_F)$. For notation simplicity we take $g(n) = g_{+}(n) = 2(|\delta \Delta| - |v_F/\ell|n)$ and from Eq. (\ref{eq:localisation}) one finds that $n\in \mathbb{N}$ is such that
\begin{align}
	n < N_{max.} = \frac{\ell |\delta \Delta|}{|v_F|} \left( 1 - \sqrt{ \left| \frac{\bar{\Delta}}{{\delta \Delta}}\right|} \right),
\end{align}
thus one can only find bound states ($N_{max.} \geq 0$) if $|\bar{\Delta}/\delta\Delta| < 1$ which corresponds to situations with gap inversion ($\Delta_1\Delta_2 < 0$).

\paragraph{Thin interface, $|\ell| < |v_F/2\delta\Delta|$.} In this situation, one can write 
\begin{align}
	g_{\sigma}(n) =  \left| \frac{2v_F}{\ell} \right| \left[ {\rm sgn}(\sigma \ell \delta\Delta/v_F) \left| \frac{\ell\delta\Delta}{v_F} \right| - n \right].
\end{align}
From this expression and from the condition for bound states (\ref{eq:localisation}), one finds that
\begin{align}
	- n > {\rm sgn}(\sigma \ell \delta\Delta/v_F) \left| \frac{\ell\delta\Delta}{v_F} \right| + \left|\frac{\ell}{2v_F} \sqrt{\bar{\Delta}\delta \Delta}\right| > {\rm sgn}(\sigma \ell \delta\Delta/v_F) \underbrace{\left| \frac{\ell\delta\Delta}{v_F} \right|}_{\in [0,\frac{1}{2}]},
\end{align}
which has the solution $n = 0$ only for the spinor $\phi_{\sigma}$ with $\sigma = - {\rm sgn}(\ell \delta \Delta/v_F)$.

The previous results show that in the smooth interface the bound states have the following spectrum, for $n \in \mathbb{N}$,
\begin{align}
	E = \pm \sqrt{ v_F^2 k_{\parallel}^2 - \frac{1}{f^2(n)}[f^2(n) - \delta \Delta^2][f^2(n) - \bar{\Delta}^2]},
\end{align}
with $f(n) = |\delta \Delta| - |v_F/\ell|n$ and $n < N_{max.} = \frac{\ell |\delta \Delta|}{|v_F|} \left( 1 - \sqrt{ \left| \frac{\bar{\Delta}}{{\delta \Delta}}\right|} \right)$. More explicitly one can write
\begin{align}\label{eq:beforesspectrumapp} 
	E &= \pm \sqrt{ v_F^2 k_{\parallel}^2 + 2 n (1 - \bar{\Delta}^2/\delta \Delta^2) \left| \frac{v_F \delta \Delta}{\ell} \right| \frac{\left( 1 - \left| \frac{v_F}{2\ell \delta \Delta} \right| n\right)\left( 1 + \frac{|v_F/\ell|}{\Delta_1} n\right)\left( 1 - \frac{|v_F/\ell|}{\Delta_2} n\right) }{\left( 1 - \left| \frac{v_F}{\ell \delta \Delta} \right| n\right)^2}}\\
	&\approx \pm \sqrt{ v_F^2 k_{\parallel}^2 + 2 n (1 - \bar{\Delta}^2/\delta \Delta^2) \left| \frac{v_F \delta \Delta}{\ell} \right|} + o\left( \frac{v_F}{\ell} \right)
	\label{eq:finalsspectrumapp}
\end{align}
which correspond to the limit of a large interface, $\ell \gg v_F/\delta \Delta$.

\subsection{b. Surface states for the linearized potential}
\label{app:linearized}

In order to obtain more physical insight in the nature of the VPS, let us now focus on the limit of a linearized interface $\ell \gg \xi$, for which the length $\ell_n$ in Eq.~(3) of the main text of the lower energy states ($n \ll N$) yields $ \ell_{n} \approx \ell_{S} / \sqrt{2n}$ with  $\ell_S = \sqrt{ \ell \xi / (1-\delta^2)}$. The values for $\ell_S$ \footnote{For example, in Ref. \cite{exp4} the convenient limit $\ell_S = \sqrt{\ell\xi}$ was considered.} and $N$ \footnote{In general, the criterium for $N$ is that the band gap of surface states should be within the smallest bulk band gap.} strongly depend on the underlying interface potential. The spectrum \eqref{eq:beforesspectrumapp} is then identical to Landau bands of the Dirac equation in a uniform magnetic field with a magnetic length $\ell_S$ \eqref{eq:finalsspectrumapp}. The relation to Landau levels can be made explicit by linearizing the gap function $\Delta(z)$ around the interface with $f(z/\ell) = v_F z/\ell_S^2$ in \eqref{eq:Hsf}. Choosing $z=0$ as the position where $\Delta(z)$ changes sign, we write 
\begin{align}
	\Delta(z) \simeq  {\rm sgn}(\Delta_2 - \Delta_1) v_F z/\ell_S^2,
\end{align}
as in Ref.~\onlinecite{interpret}. The operators $\hat{c} = \ell_S\hat{a}/\sqrt{2}v_F  ,\hat{c}^{\dagger} = \ell_S \hat{a}^{\dagger}/ \sqrt{2}v_F $ act as ladder operators, 
$\left[ \hat{c}, \hat{c}^{\dagger} \right] =  {\rm sgn}(\Delta_2 - \Delta_1) $. Following the procedure for Landau bands [\onlinecite{Marko}], in the case  
$\Delta_2 > 0 > \Delta_1$  \footnote{In the other case $\Delta_1 > 0 > \Delta_2$, the spinor components are interchanged.}, we write the eigenstates in the 
form $\Psi_n = \left[ \alpha_{1,n} |n-1\rangle,\alpha_{2,n} |n-1\rangle,\alpha_{3,n} |n\rangle, \alpha_{4,n} |n\rangle\right]$. The eigenstates  $|n\rangle$
of the number operator $\hat{n} = \hat{c}^{\dagger}\hat{c}$ are the usual harmonic-oscillator wavefunctions, in terms of the Hermite polynomials $H_n(z)$, 
\begin{equation}
 \psi_n(z)\propto H_n(z/\ell_S)e^{-z^2/4\ell_S^2},
\end{equation}
centered at the interface (around $z=0$) with a typical localization length 
\begin{equation}
 \sqrt{2n}\ell_S = \sqrt{2n\ell\xi/(1-\delta^2)}
\end{equation}
due to their Gaussian factor. Notice that the expression of the localization length coincides with that [Eq. (3)] of the main text in the limit of a smooth interface, for
$\ell\gg \xi$. 

The spectrum and eigenstates for $n\geq 1$ are obtained by diagonalizing the Hamiltonian 
\begin{align}\label{eq:Hsfn}
    \hat{H}_{c,n} = v_F
    \left[
        \begin{array}{cccc}
            0 &  k_{+} & 0 &  \frac{\sqrt{2n}}{\ell_S}\\
             k_{-} & 0 &\frac{\sqrt{2n}}{\ell_S} & 0\\
            0 & \frac{\sqrt{2n}}{\ell_S} & 0 & - k_-\\
             \frac{\sqrt{2n}}{\ell_S} & 0 & - k_+ & 0
        \end{array}
    \right] . 
\end{align}
 Moreover, the $n = 0$ state is special in that it is chiral with $\Psi_0 = \left[0,0,\alpha |0\rangle, \beta |0\rangle \right]$ and the Hamiltonian
acting on the $(\alpha,\beta)$ coefficients is  
$\hat{H}_{c,0}  = v_F ( k_y \hat{\sigma}_x - k_x \hat{\sigma}_y)\otimes \hat{P}_{{\rm sgn}(\Delta_2 - \Delta_1)}$ 
where $\hat{P}_{\sigma} = [\hat{\tau}_z -  \sigma\hat{\mathbbm{1}}]/2$ is a projection operator on the chiral $|\sigma\rangle = |\pm\rangle$-states.

\section{II. Surface states in an electric field}

\subsection{a. Lorentz boost}

We assume a $z$-dependent chemical potential $\mu(z) = \frac{1}{2} (\mu_2-\mu_1) f(z/\ell)$  in Eq. \eqref{eq:Hsf} which has the same 
profile $f(z/\ell)$ than the gap $\Delta(z)$ with $f(\pm \infty) = \pm 1$.
Performing a Lorentz boost \cite{exp4,wsmss} on Eq. \eqref{eq:Hsf} with $\mu(z)$, one finds  $ |\tilde{\Psi}\rangle = \mathcal{N} e^{-\eta \hat{\mathbbm{1}} \otimes \hat{\tau}_z /2}|\Psi\rangle$
in the new frame of reference. The Schr\"odinger equation then 
becomes $\hat{H}_c'|\tilde{\Psi}\rangle = \varepsilon |\tilde{\Psi}\rangle$ for 
$\tanh(\eta) \equiv \beta = - (\mu_2-\mu_1)/(\Delta_2 - \Delta_1) \in [-1,1]$, with
\begin{align}
    \hat{H}_c' = - \frac{1}{2}(\mu_2-\mu_1) \delta \hat{\mathbbm{1}} + \hat{H}_c(v_F', \xi', \delta', \ell)
    \label{eq:Hcl}
\end{align}
and $\hat{H}_c$ defined in Eq. \eqref{eq:Hsf} with $v_F' = \sqrt{1 - \beta^2} v_F$, $\xi' = \xi/\sqrt{1-\beta^2}$ and
\begin{align}\label{eq:effdelta}
    \delta' = \frac{1}{1-\beta^2} \left[ \delta + \frac{\varepsilon (\mu_2-\mu_1)/2 }{(v_F/\xi)^2} \right].
\end{align}
The surface states spectrum with and without a chemical potential drop are thus related by renormalized $v_F$, $\xi$ and $\delta$, and by a shift in the spectrum of $\mu_S = - \frac12 (\mu_2 - \mu_1)\delta$. This shift is used in ARPES measurements \cite{exp2,exp3} for estimating the electrostatic band bending within the hypothesis $\delta = 1$.

\subsection{b. Case of a linearized potential}

Much intuition can be also gained by considering a linearized interface $\ell \gg \xi$ ({\it i.e.} a uniform electric field)  corresponding to a spectrum  
 \begin{align}
    \varepsilon_{n,\pm} = - \frac12 (\mu_2 - \mu_1) \delta \pm v_F' \sqrt{k_{\parallel}^2 + 2(1-\beta^2)^{1/2}n/\ell_S^2 },
\end{align} 
where $\ell_S = \sqrt{\ell\xi}$ is independent of $\delta$ [\onlinecite{exp4}]. 
We recover the flattening of surface states band dispersion with $v_F' = \sqrt{1-\beta^2} v_F$ \cite{volk,interpret,vfelec1,vfelec2}, 
the reduction of the band gap of the VPS with $(v_F/\ell_n)' = (1-\beta^2)^{3/4} v_F/\ell_n$ \cite{lor1,lor2,lor3,exp4} and, moreover we identify the surface chemical-potential as $\mu_S = - \frac12 (\mu_2-\mu_1) \delta$, which corresponds to the value of $\mu_S = \mu(z_0)$ at the position $z_0$ where gap vanishes, $\Delta(z_0) = 0$.
This surface chemical potential $\mu_S$ naturally depends on the gap asymmetry ($\delta\neq 0$) :  the surface states are restricted within the smallest band gap on each side of the THJ, corresponding to a chemical potential drop smaller than the critical voltage $ |\mu_2-\mu_1|< eV_c$.

Note that the chemical doping of the pn-junction is $\mu_c$. In the case $|\Delta_1| < |\Delta_2|$, with the convention of opposite chemical potentials in each bulk semiconductor, its value depends on $\mu_2-\mu_1$: ($i$) for $\mu_2-\mu_1 < -(\Delta_2+\Delta_1)$, $\mu_{c,1} = \frac12(\Delta_2+\Delta_1)$, ($ii$) for $-(\Delta_2+\Delta_1) < \mu_2-\mu_1 < \Delta_2+\Delta_1$, $\mu_{c,2} = -\frac12(\mu_2-\mu_1)$ and, ($iii$) for $\mu_2 - \mu_1 > \Delta_2+\Delta_1$, $\mu_{c,3} = -\frac12 (\Delta_2+\Delta_1)$. The surface doping is $\mu_{c,s} = \mu_c-\mu_S$ and with our model $|\mu_{c,s}| < |\delta \Delta_1|$. 

\section{III. Stability of the chiral surface state}
\label{app:acasher}

The gapless surface state ($n = 0$) does not depend on $\Delta(z)$ nor on the interface width $\ell$. 
The chiral eigensolutions $\Psi_+ = (\phi_{1,+}, \phi_{2,+}, 0, 0)$ and $\Psi_- = (0, 0, \phi_{1,-}, \phi_{2,-})$ to the Hamiltonian (\ref{eq:Hsf}) 
correspond to the $n = 0$ surface state, with a spectrum $\varepsilon_{\pm} = \pm v_F |{\bf k}_{\parallel}|$. 
Within the Aharonov-Casher argument \cite{acasher,kawaraba}, only $\Psi_+$ or $\Psi_-$ is a bounded solution, as demonstrated in Refs. \cite{volk, volk1, volk2}. Indeed, the component  $\phi_{i,s}$ ($i = \uparrow\downarrow$; $s = \pm$) is a solution of
\begin{align}
	 \left[v_F \partial_z + s \Delta(z) \right] \phi_{i,s} = 0,
\end{align}
for which the long-range behavior is 
(i) $\phi_{i,s} \sim e^{ - s \lambda_1 z}$ with $\lambda_1 = \Delta_1 /v_F$ for $z < 0$ and 
(ii) $\phi_{i,s} \sim e^{- s \lambda_2 z}$ with $\lambda_2 = \Delta_2 /v_F$ for $z > 0$.
In the case of an infinite-sized sample, these solutions decay only if $s \lambda_2 > 0 > s \lambda_1$ 
which corresponds to $s = {\rm sgn}(\Delta_2 - \Delta_1)$ and $ \Delta_1 \Delta_2 < 0 $. Thus, the $n = 0$ mode exists as soon as there is band 
inversion and its chirality is given by ${\rm sgn}(\Delta_2 - \Delta_1) $.

\section{IV. Interpretation of ARPES data in terms of a topological heterojunction}

For Bi$_2$Se$_3$, one finds $v_F = 2.3$ eV\,\AA\ \cite{vf1} to $v_F = 3\dots 5$ eV\,\AA\ \cite{vf2} and $2\Delta = 350$ meV \cite{exp1}, and thus 
$\xi \approx v_F/\Delta = 6.5\dots 23$ \AA. 
In an oxidizing atmosphere, an oxide layer forms and we estimate the size of the interface as its depth $\ell \approx 1\dots 2$ nm \cite{oxlayer}. 
We thus expect $N \approx \ell/\xi = 1\dots 3$ VPS, as observed in \cite{exp2,exp3}. 
Moreover, we thus find $\ell_S = \sqrt{\ell\xi} \approx 8\dots 20$\AA ~ and thus an order of magnitude for the VPS band gaps $\Delta_{VPS} \approx v_F/\ell_S = 100\dots 600$ meV which is in reasonable agreement with the results found in \cite{exp2,exp3}.

From \cite{exp2}, we can extract the following band gaps: ($i$) $\Delta_{+,n=1} \approx 330$ meV and $\Delta_{+,n=2} \approx 450$ meV for the electron-like VPS and ($ii$) $\Delta_{-,n=1} \approx -330$ meV and $\Delta_{-,n=2} \approx -400$ meV for the hole-like VPS. 
The ratios of the $n=1$ and $n=2$ VPS band gaps from the Dirac point are expected to be $\sqrt{2} \approx 1.4$, within the linear-gap approximation, and here we find $\Delta_{+,2}/\Delta_{+,1} \approx 1.36$ and $\Delta_{-,2}/\Delta_{-,1} \approx 1.2$. 
We read these bands gaps from the extremal surface potential in \cite{exp2} which we identify as $V_s = \pm 200$ meV by setting $V_s = 0$ as the potential were the VPS are the furthest to the Dirac point. 
In our theory we expect the \emph{same} band gaps for the electron-like and the hole-like VPS for opposite band bending since all quantities depend on $\beta^2 \sim V_s^2$. 
The fact that the experimentally observed gaps are roughly the same is a strong indication of the topological origin of the surface states, in agreement with our theoretical model, and indicates the absence of relevant band bending. 

The breakdown voltage for a THJ involving Bi$_2$Se$_3$ is $eV_c > 2\Delta = 350$meV. 
We thus expect that $\beta = V/V_c$ introduced in the main text is $\beta < \beta_{max} = V_{s}/V_c = 0.56$. Thus, the renormalisation in \cite{exp2} of the Fermi velocity is at most $v_F^{\prime}/v_F = \sqrt{1-\beta^2} = 0.82$ and that of the band gap is at most $\Delta^{\prime}/\Delta = (1-\beta^2)^{3/4} = 0.75$. 
It is hard to tell from figures in \cite{exp2,exp3} if these quantities are indeed renormalized within the reading precision. 

These points show that a quantitative analysis of the ARPES measurements on oxidized Bi$_2$Se$_3$ may provide insights in the band inversion surface states and help 
identify the breakdown voltage. 

\bibliographystyle{apsrev4-1}
\bibliography{bibliography_SurfaceState}

\end{widetext}

\end{document}